\documentclass[namedreferences]{SolarPhysics}
\usepackage[optionalrh]{spr-sola-addons} 
\usepackage{graphicx}                    
\usepackage{color}                       
\usepackage{url}                         

\begin{document}

\begin{article}

\begin{opening}

\title{On Signatures of Twisted Magnetic Flux Tube Emergence}

%
\author{S.~\surname{Vargas Dom\'{i}nguez}$^{1}$\sep
    	D.~\surname{MacTaggart}$^{2}$\sep
	L.~\surname{Green}$^{1}$\sep      
	L.~\surname{van Driel-Gesztelyi}$^{1,3,4,5}$\sep
	A.W.~\surname{Hood}$^{6}$
       }

%
\runningauthor{Vargas Dom\'{\i}nguez et al.}
\runningtitle{On Signatures of Twisted Magnetic Flux Tube Emergence}

%
  \institute{$^{1}$ UCL-Mullard Space Science Laboratory, Holmbury St Mary, Dorking, Surrey, RH5 6NT, England \\ 
	     $^{2}$ Niels Bohr International Academy, Niels Bohr Institute, Blegdamsvej 17, 2100 Copenhagen, Denmark\\
							email: \url{mactag@nbi.dk} \\
	     $^{3}$ Centre for Plasma Astrophysics,  K.U. Leuven, Celestijnenlaan 200B, 3001 Leuven, Belgium\\
	     $^{4}$ Observatoire de Paris, LESIA, FRE 2461 (CNRS), Meudon Cedex, France\\
	     $^{5}$ Konkoly Obervatory of the Hungarian Academy of Sciences, Budapest, Hungary\\
 		  $^{6}$ School of Mathematics and Statistics, University of St Andrews, North Haugh, St Andrews, Fife, KY16 9SS, Scotland
             }

\begin{abstract}
Recent studies of NOAA active region 10953, by Okamoto {\it et al.} ({\it Astrophys. J. Lett.} {\bf 673}, 215, 2008; {\it Astrophys. J.} {\bf 697}, 913, 2009), have interpreted photospheric observations of changing widths of the polarities and reversal of the horizontal magnetic field component as signatures
of the emergence of a twisted flux tube within the active region and along its internal polarity inversion line (PIL). A filament is observed along the PIL and the active region is assumed to have an arcade structure. To investigate this scenario, MacTaggart and Hood ({\it Astrophys. J. Lett.} {\bf 716}, 219, 2010) constructed a dynamic flux emergence model of a twisted cylinder emerging into an overlying arcade.  The photospheric 
signatures observed by Okamoto {\it et al.} (2008, 2009) are present in the model although their underlying physical mechanisms differ.  The model also produces two additional signatures that can be verified by the observations.  The first is an increase 
in the unsigned magnetic flux in the photosphere at either side of the PIL.  
The second is the behaviour of characteristic photospheric flow profiles associated with twisted flux tube emergence.  We look for these two signatures in AR 10953 and find negative
results for the emergence of a twisted flux tube along the PIL.  Instead, we interpret the 
photospheric behaviour along the PIL to be indicative of photospheric magnetic 
cancellation driven by flows from the dominant sunspot.  Although we argue against flux emergence within this particular
region, the work demonstrates the important relationship between theory and 
observations for the successful discovery and interpretation of signatures of 
flux emergence.
\end{abstract}

%

\end{opening}

%
\section{Introduction}
The nature of the emergence of twisted magnetic flux tubes from the solar interior to the atmosphere is a current topic of debate.  Observers (Okamoto {\it et al.}, 2008, 2009; Lites, 2009; Lites {\it et al.}, 2010) have interpreted  certain observational signatures to be evidence for the \emph{bodily} emergence of twisted flux tubes within active regions to form and maintain filaments.  By this we mean that the flux ropes preserve their general structure and rise into the atmosphere as `solid cylinders'.  Theoretical investigations of twisted flux tube emergence are generally at odds with this picture.  For the past decade, simulations of the emergence of twisted cylinders have shown that the axis of the flux rope cannot rise to the corona.  This is true of simulations that use buoyancy as the rise mechanism (Fan, 2001; Archontis {\it et al.}, 2004; Arber {\it et al.}, 2007) and of those that use an initial velocity perturbation (Magara and Longcope, 2003).  It was found by Hood {\it et al.} (2009) and MacTaggart and Hood (2009) that changing the geometry of the flux rope from cylindrical to toroidal allows for the axis to rise to the corona.  This is because the toroidal shape allows plasma to drain efficiently.  In the cylindrical model, it collects in dips and this pins down the rope's axis.

In both the cylinder and torus models, however, the magnetic field from the emerging rope expands greatly when in the atmosphere. This is due to the rapid drop in the external pressure with height in the atmosphere.  The magnetic field here no longer resembles a `solid' flux rope.  This does not preclude the existence of twisted flux tubes in the solar atmosphere, however.  Both the cylindrical (Manchester {\it et al.}, 2004; Archontis and T\"{o}r\"{o}k, 2008; Fan, 2009) and toroidal (Hood {\it et al.}, 2009; MacTaggart and Hood, 2009) models form atmospheric flux ropes, {\it in situ}, through reconnection in a sheared arcade.  An observational signature of this could be photospheric flux cancellation (van Ballegooijen and Martens, 1989) and this has been investigated recently by Green {\it et al.} (2011).

\begin{figure}
\centering
\includegraphics[width=1.\linewidth]{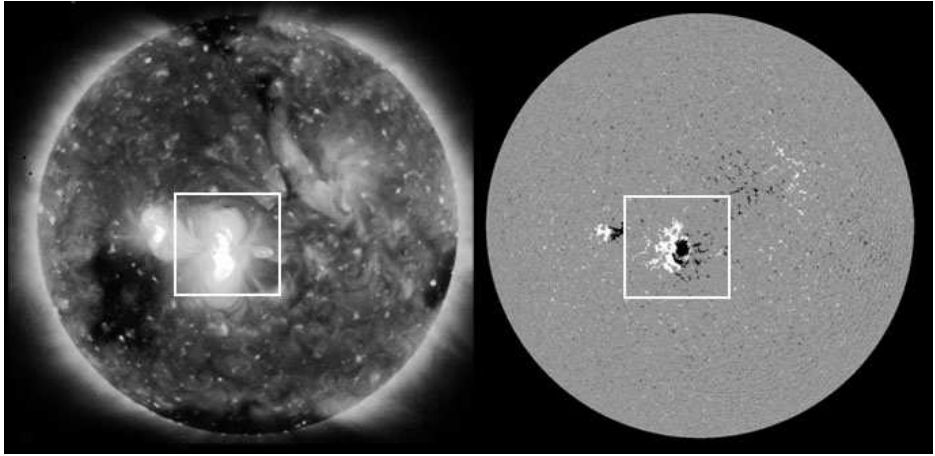}
\caption{Solar configuration observed on 1 May, 2007 using satellite telescopes:  \emph{Hinode}/XRT (left) and SOHO/MDI (right). The region analyzed in this work is framed in a white box.}
\label{fulldisk}
\end{figure}

In this paper, we address the problem of twisted flux tube emergence using a combined theoretical and observational approach.  To do this, we shall focus on NOAA AR 10953.  This has been the subject of much research in recent years (Okamoto {\it et al.}, 2008, 2009; Su {\it et al.}, 2009; Canou and Amari,
2010).  It is a complex region that includes a dominant sunspot and a filament which lies along the polarity inversion line (PIL) of a magnetic arcade, see Figure \ref{fulldisk}. Okamoto {\it et al.} (2008, 2009; hereafter, O1 and O2, respectively) studied this region with the Solar Optical Telescope (SOT) on board the \emph{Hinode} satellite (Kosugi {\it et al.}, 2007).  They argued that during the period of observation, 28 April to 9 May, 2007, a twisted flux tube emerged along the PIL beneath the filament.  This argument is based on two photospheric signatures.  The first is the widening and narrowing of the opposite polarites of the magnetic arcade.  This is interpreted as the flux rope pushing the arcade field aside as it rises through the base of the photosphere.  The second signature is the change in orientation of horizontal magnetic fields at the PIL from a normal-polarity configuration to an inverse one.  {\it i.e.} vectors pointing across the PIL in one direction change to point across the PIL in the other direction.  This is a natural consequence of a flux rope rising through the base of the photosphere (or the plane of the magnetogram).

Based on this bodily flux emergence interpretation, \inlinecite{dmac10} (the companion paper to this work, hereafter called MH10) constructed a dynamic flux emergence model of a long twisted flux rope emerging into an overlying magnetic arcade.  The initial equilibrium for the model is a plane-parallel stratified atmosphere consisting of a solar interior (marginally stable to convection), a photosphere/chromosphere, a transition region and a corona.  An equilibrium magnetic arcade is present in the atmosphere (photosphere and above) and is rooted in the solar interior.  The solar interior is field-free apart from the legs of the arcade.  A buoyant twisted flux cylinder is placed in the solar interior and rises to the base of the photosphere where it emerges into the overlying arcade.  Full details of parameter values can, of course, be found in MH10.  The two signatures, described above, are present in the model.  The interpretation of these signatures differ, however, from those proposed by O1 and O2.  For the first signature, described as the `sliding doors' effect by MH10, the broadening and narrowing of the polarities is not associated with the motion of the footpoints of the overlying arcade, but is instead related to the flux rope itself.  When the flux rope rises to the base of the photosphere it makes its own magnetic imprint on the magnetograms.  Since the flux rope cannot rise in the photospheric layer due to buoyancy alone, it begins to expand laterally, pushing into the opposite polarities of the arcade.  On magnetograms produced from the simulation, the opposite polarities of the flux rope and the overlying arcade appear to merge.  It is then that the `sliding doors' effect is found in the model.  MH10 explain this effect as being due to different stages in the emergence process - the broadening due to lateral expansion at the photosphere and the narrowing due to emergence into the high atmosphere via the magnetic buoyancy instability.

The mechanism, in the model, for the second signature is similar to that described by O1 and O2. {\it i.e.} it represents the axis of the flux rope rising above the plane of the magnetogram.  The difference between the model and the interpretation of the observations, however, is that the axis of the rope only rises approximately one photospheric pressure scale height ($\approx$ 170~km) in the model and does not reach coronal heights.  The top of the flux rope, in the model, does, however, reach the corona due to the magnetic buoyancy instability.

Although the model of MH10 does not support the bodily emergence of a twisted flux rope or the exact explanations of the signatures given by O1 and O2, it does produce those signatures for the emergence of a twisted flux tube. It may be argued, therefore, that this supports the idea that a twisted flux tube is emerging along the PIL in AR 10952.  These signatures, however, are not the only indicators of flux emergence.  In the model there are two other signatures that can be verified in observations.  These other signatures are the increase in unsigned flux at the photosphere due to the presence of the flux rope and the photospheric flow profiles produced along the PIL due to the emergence of the flux rope.  Here we argue that the signatures of O1 and O2 are neccessary but not sufficient indicators of twisted flux tube emergence and that other signatures must also be considered.  In the following sections we shall consider the two extra signatures.  We examine what form they take in the model and look for these in the observations.  The paper concludes with a discussion of the observational results and the theory of bodily and non-bodily emergence.

\section{First Signature: Unsigned Flux}
\begin{figure} 
 \centerline{\includegraphics[scale=0.45]{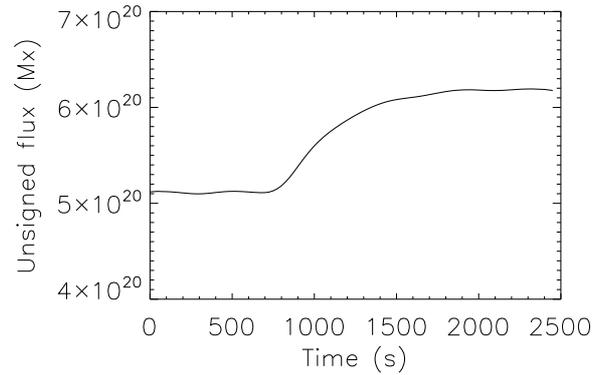}}
 \caption{Modelling result: the evolution of the unsigned vertical flux from negative polarites for the duration of the simulation.}\label{model_flux}
 \end{figure}

If a flux rope rises to the photosphere, irrespective of whether the subsequent emergence is bodily or not, there must be an initial increase in the unsigned vertical flux in the photosphere,

\[
 \Phi = \int\int|B_z|\,{\rm d}x\,{\rm d}y,
\]
where the integral is evaluated in the plane of the magnetogram.  For the model, the evolution of the vertical flux for negative polarities through time is displayed in Figure \ref{model_flux}.  The reason for only considering negative polarities is to facilitate a better comparison with observations (which is discussed later).  Due to computational constraints ({\it e.g.} run times, data storage, etc.), the dimensions of the flux systems in the model are smaller than those of AR 10953.  This is a common feature in most current dynamic simulations of the emergence of magnetic flux ropes.  Hence, the absolute values on the axes of Figure \ref{model_flux} cannot be compared directly with the values from observations. However, it is the trend of the curve showing how the amount of unsigned flux changes in time that is important.  Similar curves have been found from flux calculations for other active regions (Wang and Zirin, 1992).  It is now possible to calculate the flux as a function of time from the observations and compare the trend to that from the model.

\begin{figure} 
 \centerline{\includegraphics[scale=0.5]{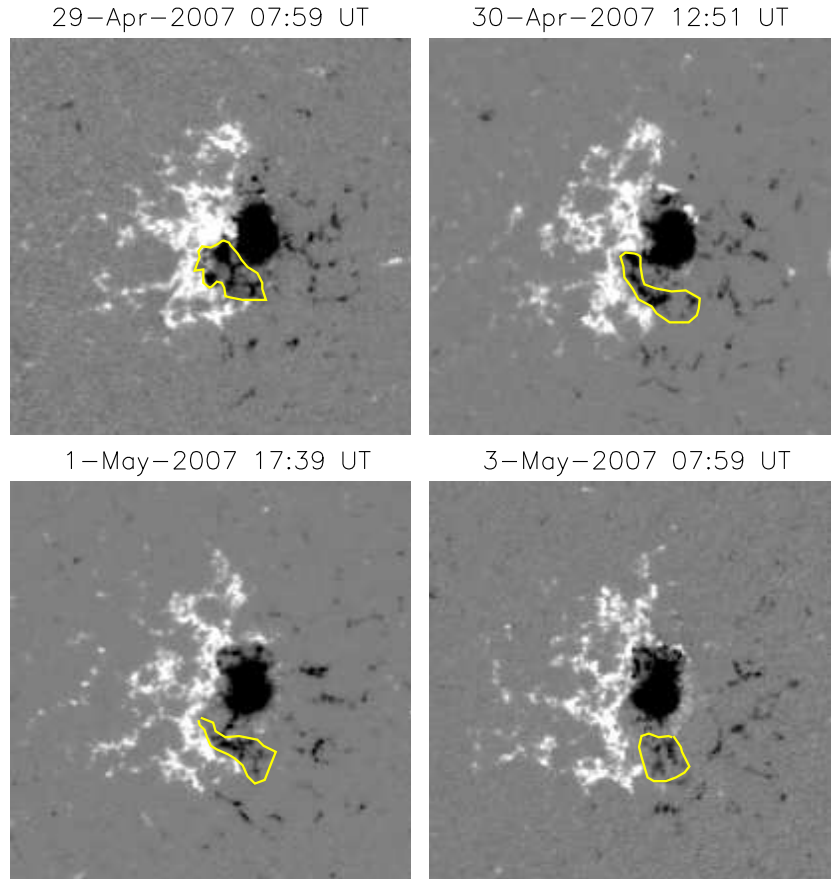}}
 \caption{The contour (yellow border on MDI magnetograms) at four different times, within which the negative flux is calculated.}\label{flux_MDI}
 \end{figure}

The evolution of the flux along the PIL is studied using SOHO/MDI magnetograms 
from 29 April to 3 May (the duration of the proposed emergence). MDI
measures the line of sight magnetic field in the mid-photosphere (Scherrer {\it et al.}, 1995) and the data have a cadence of approximately 96 min. The dataset used includes 
a mix of single images with a 30~s exposure time and an average of five such
images. The noise level per pixel is 20~G (gauss) and 9~G respectively.  
A suitable contour must be chosen around the flux region (which can deform with
time)  in order to compute the flux evolution.  One has to be careful that no 
flux leaves or enters the region defined by the contour.  For this reason, we 
choose to follow the negative polarity element which runs along the PIL where 
O1 and O2 propose that flux emergence occurs.  The evolution of this flux
region can be followed in its entirety. The positive polarity is too 
extended to be able to isolate the fragment at the PIL involved in the proposed
flux emergence event. Figure \ref{flux_MDI} displays MDI magnetograms of the active region for four different times during the proposed flux rope emergence phase.  The MDI data are scaled between $\pm$ 500 G. They have been corrected for area foreshortening that occurs away from the central meridian and the radial field component has been estimated using the IDL Solar Software routine \texttt{zradialise}.  Each panel is $400"~\times~400"$ and the yellow contour shows the region within which the negative flux is computed.  The contour is defined by eye.

\begin{figure} 
 \centerline{\includegraphics[scale=0.5]{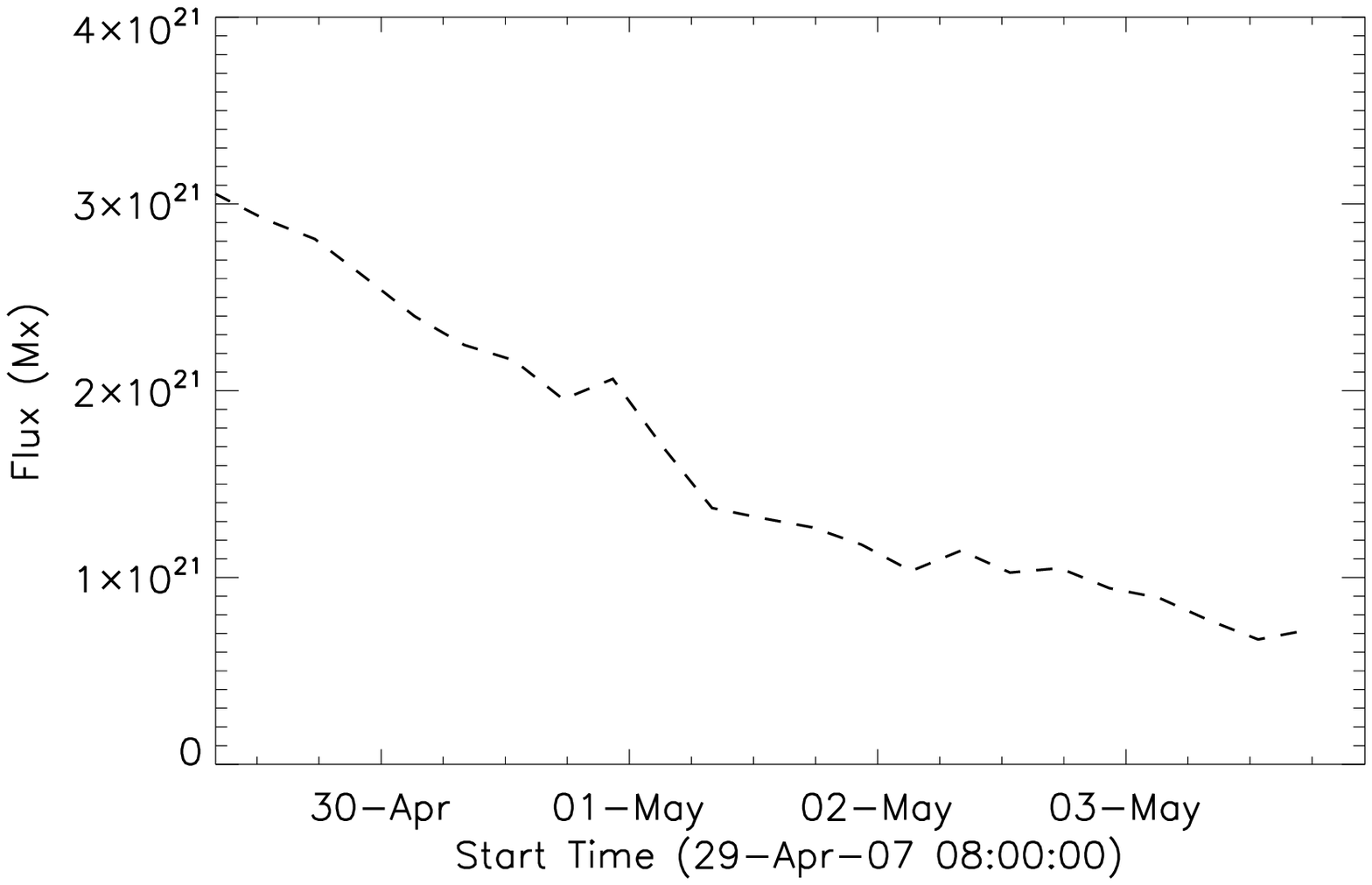}}
 \caption{The negative vertical flux, calculated within the yellow contour, as a function of time.}\label{flux_rate}
 \end{figure}

Many negative polarity fragments peel away from the negative sunspot on the right so it is important that the contour deforms to avoid these.  We calculate the flux contained within the contour and the evolution in time is shown in Figure \ref{flux_rate}.

The major trend of this figure is a decrease of negative flux with time, the opposite to the simulation prediction (see Figure \ref{model_flux}).  This,
along with the disappearance of negative fragments which are butted up against positive flux at the PIL, suggest that flux cancellation is occuring rather than flux  emergence.  The work of Su {\it et al.} (2009), who model the overlying filament using a non-linear force-free model, agrees with this result.  They speculate that the flare reconnection associated with the filament is related to cancellation rather than emergence at the PIL.

Calculating the negative flux for the \emph{entire} active region, we confirm that it is in the decay phase. On the 2nd of May the AR had 1.8$\times 10^{22}$~Mx (maxwell), a reduction from 2.3$\times 10^{22}$~Mx on the 29th of April.  The AR flux, due to cancellations, decreased on average by about 9$\%$ per day.  Similar values apply to the total positive flux.

\section{Second Signature: Photospheric Flow Profiles}
\begin{figure}
\includegraphics[angle=-90,width=.6\linewidth]{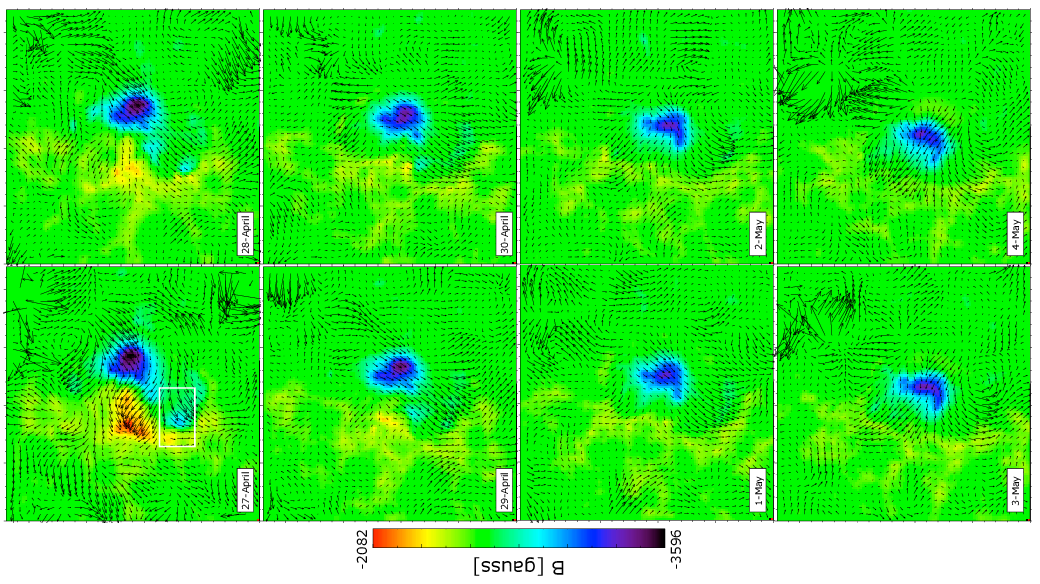}
 \caption{Daily flow maps (tracking window of FWHM $20"$) from MDI
data. Every map is computed over 15-images time series.
Arrows show the direction of horizontal flows in the FOV (~$600" \times 600"$)
The background in every case represents, in false-colour, the
average image of the corresponding day.}\label{flows}
 \end{figure}

The first signature implies that cancellation is occuring at the PIL due to magnetic fragments peeling off the negative sunspot.  In order to quantify the horizontal velocities from the observations, we apply the local correlation tracking (LCT) technique (November and Simon, 1988). This technique enables us to compute the proper motion of magnetic elements over the MDI time series of magnetograms by using a Gaussian tracking window. Maps for every single-day of the observation (15 images per day) from 27th April to the 4th May were independently computed for the active region, with a tracking window of FWHM 20". Figure \ref{flows} shows the daily flow maps for the period of interest where the background represents, in false-colour, the corresponding average magnetic field image for every day.

Upon inspection, the horizontal velocities appear to be dominated by flows from the dominant sunspot.  Since active regions have their own displacement while embedded in the granulation ({\it e.g} due to differential rotation and intrinsic motions) we aligned the time series to keep the sunspot as stationary as possible and performed the LCT analysis once more, with a tracking window of FWHM 10". The results, however, are qualitatively the same as those in Figure \ref{flows} and so are not displayed. It should be noted here that the flux at the PIL can clearly be seen, in Figure \ref{flows}, to decrease with time over the period of observation.  

\begin{figure}
 \includegraphics[angle=-90,width=.6\linewidth]{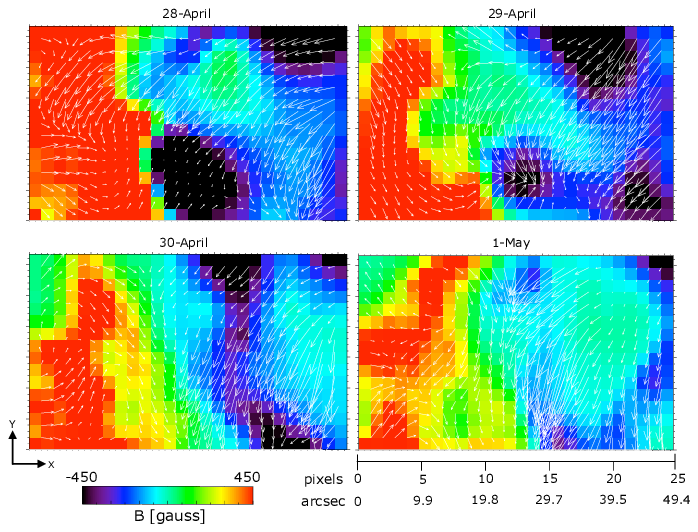}
 \caption{Flow maps at the PIL for the 24 h sets (tracking window of FWHM $8"$). The backgrounds show the corresponding average image for every day with a FOV of ~$30" \times 50"$ that corresponds to the white box in the upper left panel in Figure~\ref{flows} (clipped to -450 and 450 G).}\label{zoom_flows}
 \end{figure}

Focussing on the PIL (white box in Figure \ref{flows}), we performed LCT on a smaller tracking window of $8"$ and over 24 h and 8 h sets.  Figure \ref{zoom_flows} shows the horizontal flows for the 24 h sets on four different days.  These days cover the period of the proposed emergence.  

\begin{figure}
 \includegraphics[angle=90,width=1.\linewidth]{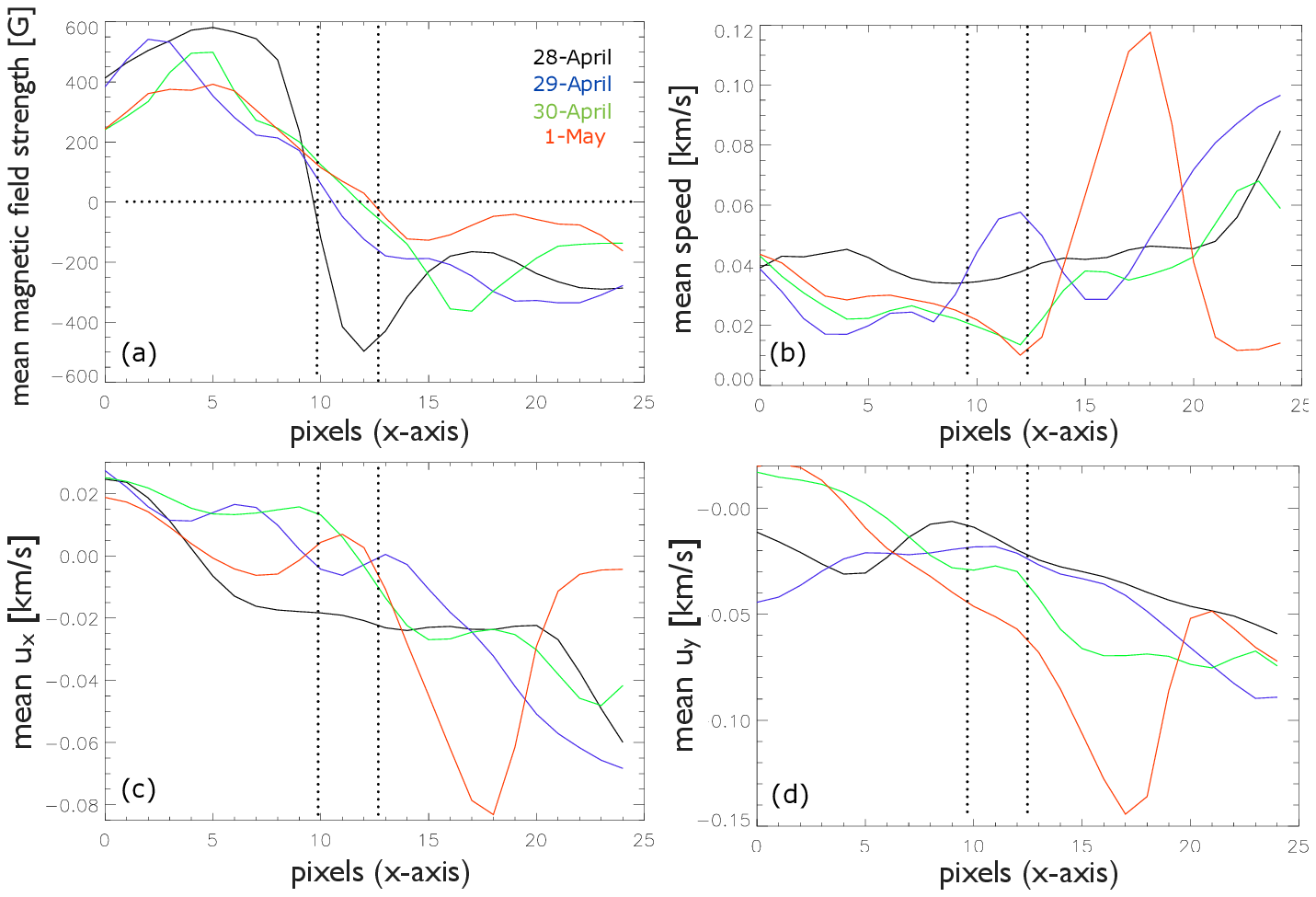}
 \caption{Profiles computed over the images in Figure~\ref{zoom_flows} by using the mean values along columns for:  (a) magnetic field strength, (b) horizontal speed, (c) $x$-component of horizontal velocity, (d) $y$-component of horizontal velocity.
Vertical lines in the top left panel indicate the region where the PIL is located during the consecutive days from 28 April to 1 May as displayed by the corresponding colour lines.
 }\label{profiles}
 \end{figure}

To aid the analysis of the horizontal flows, we compute the mean profiles of the magnetic field and horizontal speeds.  The values for each pixel are averaged over every column from Figure \ref{zoom_flows}.  These averages are combined to produce mean profiles as a function of the horizontal distance $x$. Figure \ref{profiles} displays the mean trends of the magnetic field strength, the horizontal speed $\left(u=\sqrt{u_x^2+u_y^2}\right)$, $u_x$ and $u_y$, for the period of 28 April to 1 May.

Figure \ref{profiles}~(a) shows the mean magnetic field strength, for the four days, and highlights the positions of the PIL.  It shows that the mean flux is decreasing with time, again linking with the first signature.  Figure \ref{profiles}~(b) displays the mean speed profiles.  The speeds grow as one moves past the PIL to the right.  This indicates the dynamical dominance of the right-hand sunspot. Figure \ref{profiles}~(c) displays the mean $u_x$ profiles.  
There are two general features of these mean profiles which are divided by the PIL locations.  To the right of the PIL locations, $u_x$ is negative and increases in magnitude with time.  To the left of the PIL positions, $u_x$ is either negative or positive but with magnitudes much less than values to the right of the PIL locations.  These flows represent the magnetic fragments drifting from the dominant sunspot to the PIL.  In Figure \ref{profiles}~(d), all the mean profiles of $u_y$ have negative values in the vicinity of the PIL locations.  As for the $u_x$ mean profiles, the magnitudes of the $u_y$ profiles are greater to the right of the PIL locations.  These profiles represent magnetic fragments moving downwards from the dominant sunspot.  The velocities of the LCT analysis clearly show the south-western flow of negative magnetic fragments from the dominant sunspot towards the PIL.  Here, they cancel with positive polarities, as shown by the first signature.

\begin{figure} 
\centerline{\includegraphics[width=0.5\textwidth,clip=]{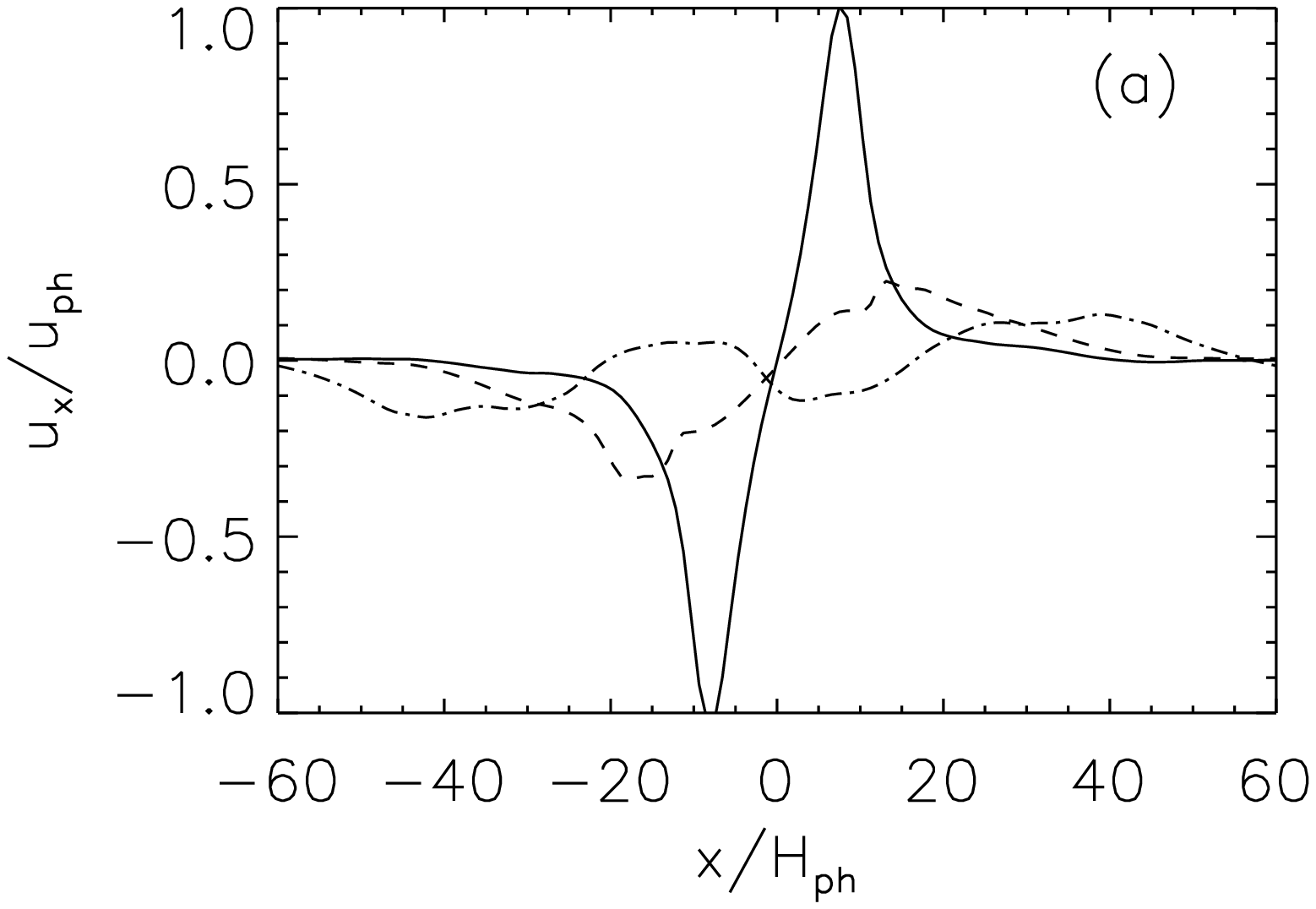}
\includegraphics[width=0.5\textwidth,clip=]{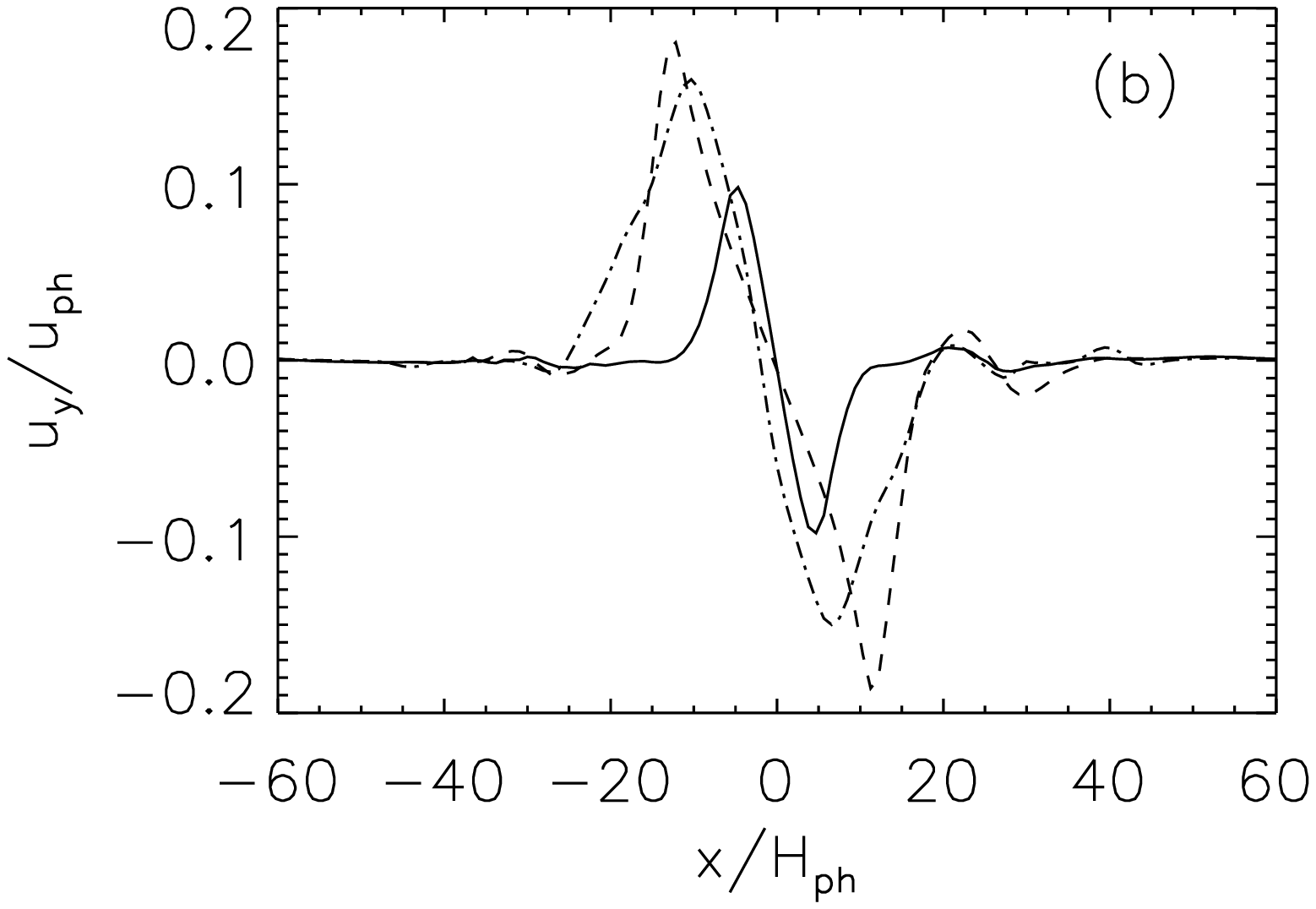}}
 \caption{Photospheric flow profiles from MH10. (a) displays $u_x$ profiles as a function of $x$ and (b) displays $u_y$ profiles as a function of $x$ for three different times. The PIL is located at $x=0$.  Key: $t=900$~s (solid), $t=1225$~s (single-dash) and $t=2200$~s (dot-dash).}\label{shear}
 \end{figure}

The flow profiles from the LCT analysis give no clear indications of a twisted flux tube emerging at the PIL. Figure \ref{shear} shows the horizontal photospheric flow profiles from the simulation at three different times.  The full details of the emerging flux tube at the photosphere can be found in MH10.  Since the magnetic fields follow the velocity pattern (as $\beta \approx 1$), one would expect to find similar flows to those in Figure \ref{shear} in the LCT if a flux rope was emerging.  Figure \ref{shear} (a) shows the profiles for $u_x$.  These must be considered to be mean profiles, for comparison with observations, since the model does not explicitly include turbulent convection.  Initially, there is a diverging flow where the flux tube emerges upwards and expands.  This diverging flow weakens and then reverses to a converging flow at later times.  Although converging flows are found in the observations, the corresponding diverging flows, required by the emergence scenario, remain absent.  Figure \ref{shear} shows the profiles of $u_y$ at three different times (the same times as in (a)).  These profiles display the shear flows that develop throughout the emergence process (Manchester {\it et al.}, 2004; MacTaggart and Hood, 2009). The magnitude of these flows will be larger than what would be found in observations since we do not include convection in the model.  However,  Fang {\it et al.} (2010) include the effects of convective motions in their emerging flux rope simulation and still find distinct shear flows.  Since the magnetic field follows the velocity profiles ({\it i.e.} the magnetic field is also sheared), some evidence of shearing should appear in the LCT analysis if twisted flux tube emergence was taking place.  The results from the LCT analysis, however, show no signs of shearing at the PIL locations.

\section{Discussion}
Understanding signatures of flux emergence is essential for the correct identification of emerging flux ropes.  In this work, we have considered an observation that has been reported as an example of twisted flux tube emergence (O1, O2).  The argument for flux emergence is based on two signatures observed in the photosphere.  These signatures were then investigated by MH10 using a dynamic flux emergence simulation and were found to be present in the model but have different interpretations.  The model also produces other signatures that can be tested observationally.  In this work, we have investigated two of these extra signatures by comparing what the model predicts to what is present in the observations.

Although we are considering observations in the photosphere, the MDI magnetograms are likely to be at a different height to the SOT magnetograms.  Depending on what is being studied this may or may not have an impact on the results.  For example, assuming there is no significant rotation of the active region magnetic field, the second signature of O1, O2 and MH10 depends upon the height of the magnetogram. {\it i.e.} if the plane of the magnetogram is always above the flux rope's axis, the signature will not be detected. If there is significant rotation, {\it e.g.} due to a highly-twisted magnetic field, then this signature could be detected in a plane above the axis of the active region field (Fan, 2009) and it would not be due to any secondary emergence at the PIL.   The possible variation in magnetogram heights is not a problem for the other signatures considered in this paper.  For the first signature, if a twisted flux rope is to emerge, the field must rise through the photosphere.  Any plane in the photosphere should, therefore, show an increase in unsigned flux due to the presence of the rope.  The trends for different photospheric planes should be similar, only the times and scales will change.  The second signature should also not be affected by the choice of plane taken in the photosphere.  The Lorentz force drives shear flows at all heights in the emerging arcade and the expanding magnetic field will push plasma aside as it rises upwards (MacTaggart and Hood, 2009).

For the two signatures studied, the observations give negative results for flux emergence along the PIL of AR 10953, based on the model of MH10.  Instead they point towards magnetic cancellation at the PIL. In looking for the first signature, we found a decrease in the unsigned flux at the PIL rather than the expected increase.  As this is independent of the emergence model, this evidence strongly suggests that flux emergence does not take place at the PIL.  The second signature deals with characteristic photospheric flows, associated with twisted flux tube emergence at the PIL, derived from MH10. The LCT maps of the region, however, mainly reveal flows of magnetic fragments travelling from the dominant sunspot towards the PIL. There is no clear trace of the kinds of flows found in the simulation.  

There are two points to note about this comparison.  The first is that the simulation produces fluid velocities and these are not strictly the same as those of the moving magnetic features.  However, for emergence to occur, the plasma beta must be of order unity in the photosphere.  Hence, the fluid and magnetic velocities should correlate closely and not be independent.  The second point is that since the model of MH10 does not include turbulent convection, the flow profiles must be considered to be averages.  Also, magnitudes of the speeds from the model of MH10 will be larger and persist for longer than if turbulent convection was included.  However, for the emergence of a long coherent rope, as proposed by O1 and O2, such flow profiles must be present, albeit slightly weaker, since they follow from the basic physics of the rising rope. As stated above, our investigation of the observations does not reveal the flow profiles, predicted by the model, at the PIL. Instead of emergence, the velocity maps indicate the dynamical domiance of the right-hand sunspot.  Fragments peel off from this and move towards the PIL due to the moat flow (Vargas Dom\'{i}nguez {\it et al.}, 2007).  This produces flow profiles that are indicative of convergence and link to the result from the first signature.  

The evidence of the first signature is difficult to refute.  The presence of shear flows, from the second signature, is often argued against, however, in relation to active region filament formation.  Some observational studies ({\it e.g.} O2, \opencite{lites09}) report that they find no evidence of robust photospheric shearing at the PIL. This leads these authors to adopt the idea of bodily emergence to explain filament formation.  The shear and convergence method \cite{vanb89} cannot be completely discarded, however.  Photospheric observations are taken from a particular plane at a particular height.  It may be the case that if the plane of the observation is low enough, coherent shear flows may be disrupted by the turbulent convection.  However, as stated earlier, shearing occurs throughout the expanded emerging magnetic field \cite{dmac09}.  In the simulation of Fang {\it et al.} (2010), they find that shear flows in the corona persist even when those in the photosphere have been washed out by convective motions.  Another reason why the shear and convergence method cannot be discarded for this set of observations is that the active region is in the decay phase.  As shown by the flux and LCT studies, the main process is cancellation driven by convergence.  Since the filament already exists, this could be the late phase of the shear and convergence method.  

The formation of atmospheric flux ropes (or active region filaments) through reconnection in a sheared arcade has been reproduced in several simulations (see Introduction). Each stage of the process has been investigated and has a clear physical explanation.  The theoretical underpinning (or viability) of bodily emergence, however, remains to be tested and should be the subject of further study.  

The findings of MH10 suggest that the signatures of O1 and O2 are not sufficient to uniquely identify an emerging flux rope.  The findings of this paper suggest that, for the region studied by O1 and O2, the emergence of a long twisted flux rope does not, in fact, take place at the PIL. Although we argue against flux emergence in this region, this work does highlight the important relationship between flux emergence simulations and observations for the correct identification of twisted flux tube emergence.

%

%

%

%
 \begin{acks}
D.M. acknowledges financial assistance from STFC. The computational work for this paper was carried out on the joint STFC and SFC (SRIF) funded cluster at the University of St Andrews. D.M. and A.W.H. acknowledge financial support
from the European Commission through the SOLAIRE Network (MTRN-CT-2006-035484).  L.G. would like to thank the Royal Society for the Dorothy Hodgkin Fellowship which supported this work.  L.v.D.G. acknowledges funding through the Hungarian Science Foundation grant OTKA K81421 and the European Community's FP7/2007 � 2013 programme through the SOTERIA Network (EU FP7 Space Science Project No. 218816).  Hinode is a Japanese mission developed and launched by ISAS/JAXA,
collaborating with NAOJ as a domestic partner, NASA and STFC (UK) as
international partners. SOHO is a project of international cooperation
between ESA and NASA. We thank the MDI consortium for their data.
 \end{acks}

%
%

\begin{thebibliography}{}
 \bibitem[\protect\citeauthoryear{Arber {\it et al.}}{2007}]{arber07}
Arber, T.D., Haynes, M., Leake, J.E.: 2007, {\it Astrophys. J.} {\bf 666}, 541.
 \bibitem[\protect\citeauthoryear{Archontis and T\"{o}r\"{o}k}{2008}]{archontis08}
Archontis, V., T\"{o}r\"{o}k, T.: 2008, {\it Astron. Astrophys.} {\bf 492}, 35.
 \bibitem[\protect\citeauthoryear{Archontis {\it et al.}}{2004}]{archontis04}
Archontis, V., Moreno-Insertis, F., Galsgaard, K., Hood, A.W., O'Shea, E.: 2004, {\it Astron. Astrophys.} {\bf 426}, 1047.
 \bibitem[\protect\citeauthoryear{Canou and Amari}{2004}]{archontis04}
Canou, A., Amari, T.: 2010, {\it Astrophys. J.} {\bf 715}, 1566.
\bibitem[\protect\citeauthoryear{Fan}{2001}]{fan01}
Fan, Y.: 2001, {\it Astrophys. J. Lett.} {\bf 554}, 111.
\bibitem[\protect\citeauthoryear{Fan}{2009}]{fan09}
Fan, Y.: 2009, {\it Astrophys. J.} {\bf 697}, 1529.
\bibitem[\protect\citeauthoryear{Fang}{2010}]{fang10}
Fang, F., Manchester, W., Abbett, W.P., van der Holst, B.: 2010, {\it Astrophys. J.} {\bf 714}, 1649.
\bibitem[\protect\citeauthoryear{Green}{2011}]{green11}
Green, L.M., Kliem, B., Wallace, A.J.: 2011, {\it Astron. Astrophys.} {\bf 526}, 2.
\bibitem[\protect\citeauthoryear{Green}{2009}]{hood09}
Hood, A.W., Archontis, V., Galsgaard, K., Moreno-Insertis, F.: 2009, {\it Astron. Astrophys.} {\bf 503},
999.
 \bibitem[\protect\citeauthoryear{Kosugi {\it et al.}}{2007}]{kosugi07}
  Kosugi, T., Matsuzaki, K., Sakao, T., Shimizu, T.,
Sone, Y., Tachikawa, {\it et al.}: 2007,
{\it Solar Phys.} {\bf 243}, 3.
\bibitem[\protect\citeauthoryear{Lites}{2009}]{lites09}
Lites, B.W.: 2009, {\it Space Sci. Rev.} {\bf 144}, 197.
\bibitem[\protect\citeauthoryear{Lites {\it et al.}}{2010}]{lites10}
Lites, B.W., Kubo, M., Berger, T., Frank, Z., Shine, R., Tarbell, T., Title, A., Okamoto, T.J.,
Otsuji, K.: 2010, {\it Astrophys. J.} {\bf 718}, 474.
\bibitem[\protect\citeauthoryear{MacTaggart and Hood}{2009}]{dmac09}
MacTaggart, D., Hood, A.W.: 2009, {\it Astron. Astrophys.} {\bf 507}, 995.
\bibitem[\protect\citeauthoryear{MacTaggart and Hood}{2010}]{dmac10}
MacTaggart, D., Hood, A.W.: 2010, {\it Astrophys. J. Lett.} {\bf 716}, 219.
\bibitem[\protect\citeauthoryear{Magara and Longcope}{2003}]{magara03}
Magara, T., Longcope, D.W.: 2003, {\it Astrophys. J.} {\bf 586}, 630.
\bibitem[\protect\citeauthoryear{Manchester {\it et al.}}{2004}]{manchester04}
Manchester, W. IV, Gombosi, T., DeZeeuw, D., Fan, Y.: 2004, {\it Astrophys. J.} {\bf 610}, 588.
\bibitem[\protect\citeauthoryear{November and Simon}{1988}]{november88}
November, L.J., Simon, G.W.: 1988, {\it Astrophys. J.} {\bf 333}, 427.
\bibitem[\protect\citeauthoryear{Okamoto {\it et al.}}{2008}]{okamoto08}
Okamoto, T.J., Tsuneta, S., Lites, B.W., Kubo, M.,
Yokoyama, T., Berger, T.E., {\it et al.}: 2008,
{\it Astrophys. J. Lett.} {\bf 673}, 215.
\bibitem[\protect\citeauthoryear{Okamoto {\it et al.}}{2009}]{okamoto09}
Okamoto, T.J., Tsuneta, S., Lites, B.W., Kubo, M.,
Yokoyama, T., Berger, T.E., {\it et al.}: 2009,
{\it Astrophys. J.} {\bf 697}, 913.
\bibitem[\protect\citeauthoryear{Scherrer {\it et al.}}{1995}]{scherrer95}
Scherrer, P.H., Bogart, R.S., Bush, R.I., Hoeksema, J.T.,
Kosovichev, A.G., Schou, J., {\it et al.}: 1995,
{\it Solar Phys.} {\bf 162}, 129.
\bibitem[\protect\citeauthoryear{Su {\it et al.}}{2009}]{su09}
Su, Y., van Ballegooijen, A., Lites, B.W., Deluca, E.E., Golub, L., Grigis, P.C., Huang, G., Ji,
H.: 2009, {\it Astrophys. J.} {\bf 691}, 105.
\bibitem[\protect\citeauthoryear{van Ballegooijen and Martens}{1989}]{vanb89}
van Ballegooijen, A.A., Martens, P.C.H.: 1989, {\it Astrophys. J.} {\bf 343}, 971.
\bibitem[\protect\citeauthoryear{Vargas Dom\'{i}nguez {\it et al.}}{2007}]{vargas07}
Vargas Dom\'{i}nguez, S., Bonet, J.A., Mart��nez Pillet, V., Katsukawa, Y., Kitakoshi, Y., Rouppe
van der Voort, L.: 2007, Astrophys. J. Lett. 660, 165.
\bibitem[\protect\citeauthoryear{Wang and Zirin}{1992}]{wang92}
Wang, H., Zirin, H.: 1992, {\it Solar Phys.} {\bf 140}, 41.
%
%
%
 \end{thebibliography}
%

\end{article} 
\end{document}